# Measurement of frequency sensitivity coefficient and evaluation of type-B uncertainty for atomic frequency standard based on statistical correlation of noise

Qian Wang[1,2], Ning Zhang[1,2], Rong Wei[1,①], and Yuzhu Wang[1]

[1]Key Laboratory for Quantum Optics, Shanghai Institute of Optics and Fine Mechanics, Chinese Academy of Sciences, Shanghai 201800, People's Republic of China

[2] Center of Materials Science and Optoelectronics Engineering, University of Chinese Academy of Sciences, Beijing 100049, People's Republic of China

## Abstract

Precise measurement of frequency sensitivity coefficients (FSCs) of some physical effects contributing to uncertainty is an important work in type-B uncertainty ($u_B$) evaluation of atomic frequency standards. We proposed a method based on statistical correlation of noise to measure FSCs and evaluate $u_B$ for atomic frequency standards, and giving a full statistical expression of FSC as $K_I = \text{cov}_A(y, x_I) / \sigma^2(x_I)$, where $y$ and $x_I$ are fractional frequency of atomic frequency standards and noise independent variable of $I_{th}$ physical effects, respectively. The confidence value and the effect of time interval have also been discussed. Except for theoretical derivation, the method also has been validated by numerical simulation and demonstration experiments. Comparing with standard method, the statistical correlation of noise method combines $u_B$ evaluation with type-A uncertainty ($u_A$), measures FSCs not by special experiments, but from the monitoring data when the device is running. It not only gives a criterion to test the standard method, but also can evaluate many effects at the same time when the device is operating. The method can be extended to other precision measurement fields

---

① weirong@siom.ac.cn

and is available to comprehend the physical effectiveness of the least square method.

Introduction

The research of atomic frequency standards （AFSs） have made great progress in recent decades. Atomic fountain frequency standards (AFFSs) have realized time unit "second" with accuracy of $10^{-16}$ [1,2], and long-term frequency stability of AFFSs has reached $10^{-17}$ magnitude [3]. Stability and uncertainty of optical frequency standards (OFSs) are at the level of $10^{-18}$ [4]. The improvement of uncertainty of AFSs mean not only measuring the error of the known physical effects with higher accuracy and suppressing noise stricter, but also evaluating and decreasing the influence of more effects neglected in the past. $u_B$ evaluation of main effects is one of the most important and difficult task in the study of AFSs. Many special experiments are employed to precisely evaluate it, such as collision frequency shift[5,6], distributed cavity phase shift[7,8,9,10], Zeeman frequency shift[1,11,12] for AFFSs, and blackbody radiation frequency shift[4,13,14,15] and optical frequency shift[16,17,18] for OFSs, etc.

The evaluation for $u_B$ has a standard method adhering to some criteria[4,19,20,21], which can be summarized as following: the total error and total $u_B$ of the AFSs are the synthesis of many independent effects, the contribution can be expressed by the first-order Taylor approximation: $y_i = \bar{y}_i + k_i x_i$, where $y_i$ is fractional frequency due to the $i^{th}$ physical effect, $\bar{y}_i$ is its average value, $x_i$ is the fluctuation of the dominant noise independent variable (NIV) of the $i^{th}$ effect, and $k_i$ is FSC of the corresponding effect. $\bar{y}_i$ and $k_i$ can be evaluated by special experiments, $x_i$ is measured by monitoring for long time. The total error and total uncertainty are expressed as $\bar{y} = \sum \bar{y}_i$ and $u_B^2 = \sum k_i^2 \sigma_{x_i}^2$, respectively, where $\sigma_{x_i}$ is the uncertainty of $x_i$. In theory, the value of $k_i$ is confirmed, corresponding to the physical law of the $i^{th}$ effect, but for specific devices similar as AFSs, $k_i$ is affected by many uncontrollable conditions. There has error between its theoretical value and actual value, so measurement of $k_i$ is necessary and plays an important role in $u_B$ evaluation. For example, magic wavelength of OFSs and collision frequency shift coefficient of AFFSs are measured even if their theoretical values are well known. In addition, it is necessary to check these values regularly when the devices are running continuously.

In this paper, a method of FSCs measuring and $u_B$ evaluating is proposed and demonstrated by theory, numerical simulation and experiment, which gives the expression of FSC based on the noise statistical correlation (NSC) of frequency output and NIVs. Different from normal evaluating method of $u_B$ which measures $k_i$ and $\sigma_{x_i}$ by respective and special experiments, NSC gives the measurements entirely by statistical methods, coming from the real-time relation of $\{x_{i,j}\}$ on $\{y_j\}$. The method combines $u_B$ evaluation with $u_A$ evaluation, gives a direct and more realistic image of correlation of $y_i$ and NIVs. NSC also give views of the evolution of FSCs dependence on integral time which is help for testing and improving the measurement of NIVs, as well as understanding the relationship of the device and its environment. NSC can evaluate multiple errors of effects in parallel at the normal operating state of the device. And it is also universal in many other precision measurement fields. In the paper, we will demonstrate the method in the order of theoretical analysis, numerical simulation and experimental verification.

Theory

The relationship between the output fractional frequency of AFS and the environmental NIVs can be expressed as:

$$y_j(\tau) = \frac{1}{\tau}\int_{(j-1)\tau}^{j\tau} y(t)dt = y_{0,j}(\tau) + \bar{y} + \sum_{i=1}^{N} k_i \cdot x_{i,j}(\tau) \qquad (1)$$

Where $\tau$ is the integral time, $N$ is the number of effects counted in $u_B$ evaluation, $y_{0,j}(\tau) = \frac{1}{\tau}\int_{(j-1)\tau}^{j\tau} y_0(t)dt$ represents the noise caused by local oscillator and quantum phase discriminator system at time interval from $(j-1)\tau$ to $j\tau$, $y_{0,j}(\tau)$ characterizes the contribution of short-term frequency stability of AFS and with $\overline{y_0(\tau)}=0$. $k_i \cdot x_{i,j}(\tau)$ is the time-dependent noise contribution of the effects, affecting both $u_B$ and $u_A$. The relations among the items in the Eq.(1) also can be shown as Fig.1, where the real-time fractional frequency of the AFS is affected by two parts, one is its own characteristics related with such as Dick effect and quantum projected noise, and the other is operating environment. The former influents short-term frequency stability (spectral broadening of dashed green line), while the average result of the latter drifts the spectral center (dotted red line VS. solid blue line), whose fluctuation results in time-dependent variation of the spectral center which affects the long-term frequency stability and ultimately $u_B$.

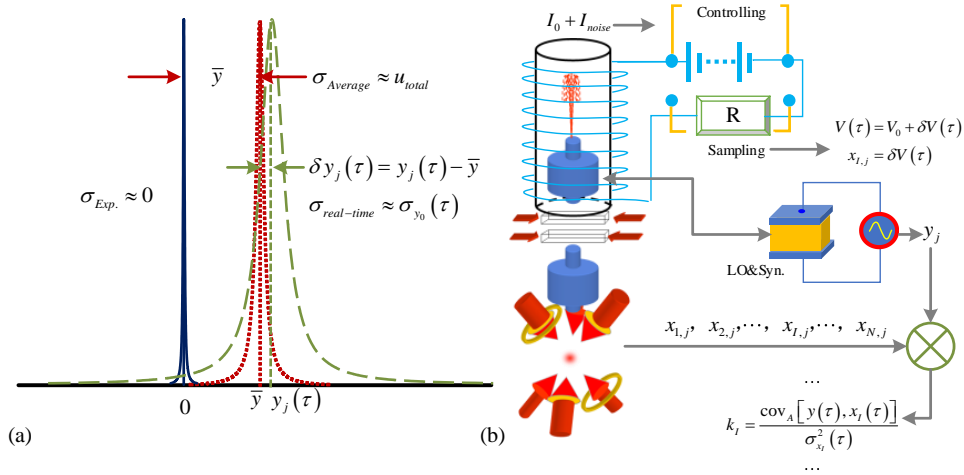

Figure 1 (color online). (a) Schematic diagrams of the noise of the AFS. (b) NSC experimental setup in an AFFS. In Fig.1 (b), the corresponding parameters are detected synchronously, then their FSCs are solved by the NSC method. An experiment is demonstrated on AFFSs by second-order Zeeman shift, where additional current noise of solenoid coil is introduced to affect the magnetic intensity of the AFFS, its FSC is evaluated from the sequences of $\{y_j\}$ and $\{\delta V_j\}$.

The total uncertainty of AFS is given by $u_{total}^2 = u_A^2 + u_B^2$, where $u_A$ is evaluated by

Allan deviation (ADEV), expressed as:

$$u_A^2 = \sigma_y^2(\tau) = \frac{1}{2(M-1)} \sum_{j=1}^{M-1} \left[ y_{j+1}(\tau) - y_j(\tau) \right]^2$$
$$\approx \sigma_{y_0}^2(\tau) + \sum_{i=1}^{N} k_i^2 \cdot \sigma_{x_i}^2(\tau) \qquad (2)$$

Here $\sigma_{y_0}^2(\tau)$ and $\sum_{i=1}^{N} k_i^2 \cdot \sigma_{x_i}^2(\tau)$ are the contributions of the quantum interrogation system and NIVs, respectively. $\sigma_{y_0}(\tau) = \frac{1}{Q} \frac{1}{SNR} \sqrt{\frac{\tau_0}{\tau}}$, where $Q$ and $SNR$ are quality factor and signal-to-noise ratio of interrogation spectrum, respectively. $\tau_0$ is the running period of the AFSs. $M = \text{int}[\tau_{total}/\tau]$ is the number of the samples in $\{y_j(\tau)\}$ sequence, where $\tau_{total}$ is total time of experiment. For the short-term stability where $\tau$ is small, $\sum_{i=1}^{N} k_i^2 \cdot \sigma_{x_i}^2(\tau)$ can be ignored for it is far less than $\sigma_{y_0}^2(\tau)$, but for the long-term stability where $\tau$ is great, $\sum_{i=1}^{N} k_i^2 \cdot \sigma_{x_i}^2(\tau)$ is main contributions for total stability. In derivation of the Eq.(2), the adopted hypothesis is that: $y_0, x_1, \cdots, x_N$ are independent of each other, so the covariance of any two of the variables is zero[19]. Extending ADEV to two-sample covariance, named Allan covariance (ACOV), expressed as:

$$\text{cov}_A(a_j, b_j, \tau) = \frac{1}{2(M-1)} \sum_{j=1}^{M-1} \left[ a_{j+1}(\tau) - a_j(\tau) \right]\left[ b_{j+1}(\tau) - b_j(\tau) \right] \approx 0 \qquad (3)$$

Where $a_j(\tau), b_j(\tau) = y_0(\tau), x_1(\tau), \cdots, x_N(\tau)$ and $a_j(\tau) \neq b_j(\tau)$. If the averages of $\Delta a_j(\tau) = a_{j+1}(\tau) - a_j(\tau)$ and $\Delta b_j(\tau) = a_{j+1}(\tau) - a_j(\tau)$ are 0, Eq.(3) is equal to $\text{cov}[\Delta a_j(\tau), \Delta b_j(\tau)]$. The covariance and ACOV of $a_j(\tau)$ and $b_j(\tau)$ tend to zero as $M$ approaches infinite, and from a statistical point of view, the smaller the value of $M$, the larger their deviation.

For the standard method, $u_B$ is expressed as $u_B^2 = \sum_{i=1}^{N} k_i^2 \cdot \sigma_{x_i}^2$, where the uncertainty of $k_i$ is ignored and $\sigma_{x_i}^2$ is written as $\sigma_{x_i}^2 = \sigma_{x_i}^2(\tau) + \sigma_{x_i-B}^2$, the sum of squares of $u_A$ and $u_B$ of $x_i$. In a few uncertainty evaluations of some effects, $\sigma_{x_i}^2(\tau)$ is often taken into account in $u_A$ rather than in $u_B$, such as collision shift evaluation of NIST-F1[4]. However in most cases, it is counted in $\sigma_{x_i}^2$ for better confidence. In the following discussion, we ignore $\sigma_{x_i-B}^2$ firstly, which means $x_i$ can be measured precisely and $\sigma_{x_i}^2$ is from the time-dependent fluctuation of $x_i$. The situation of $\sigma_{x_i-B}^2 \neq 0$ will be discussed later.

NSC focuses on the relationship between $y_j(\tau)$ and $x_{1,j}(\tau),\cdots,x_{i,j}(\tau),\cdots$. All of them are measurable quantities, and sequences of $\{y_j(\tau)\}$ and $x_{1,j}(\tau),\cdots,x_{i,j}(\tau),\cdots$ can be obtained by data monitoring when the device is running. For sample, one of $N$ effects, such as capital $I^{th}$ is discussed. In this case, the Eq.(1) can be rewritten as:

$$y_j(\tau) = y_{0,j}(\tau) + \bar{y} + \sum_{i=1}^{N} k_i \cdot x_{i,j}(\tau) = y'_{I,j}(\tau) + k_I \cdot x_{I,j}(\tau) \tag{4}$$

Here $y'_{I,j}(\tau) = y_j(\tau) - k_I \cdot x_{I,j}(\tau)$ is total fractional frequency of other effects except for the $I^{th}$ effect. If its Allan variance is $\sigma_{y'_I}^2(\tau)$, Eq.(2) can be rewritten as:

$$\sigma_y^2(\tau) = \sigma_{y'_I}^2(\tau) + k_I^2 \cdot \sigma_{x_I}^2(\tau) \tag{5}$$

Constructing a function of $Y_{I,j}(\tau)$, expressed as:

$$Y_{I,j}(\tau) = y_j(\tau) - K_I \cdot x_{I,j}(\tau) \tag{6}$$

Here $K_I$ is an introduced independent variable for FSC measuring, and its stability is given by:

$$\sigma_{Y_I}^2(\tau) = \frac{1}{2(M-1)} \sum_{j=1}^{M-1} \left[Y_{I,j+1}(\tau) - Y_{I,j}(\tau)\right]^2 \approx \sigma_{y'_I}^2(\tau) + (k_I - K_I)^2 \cdot \sigma_{x_I}^2(\tau) \tag{7}$$

we can obtain $K_I = k_I$ by solving the equation of $\frac{\partial \sigma_{Y_I}^2(\tau)}{\partial K_I} = 0$, where $\sigma_{Y_I}^2(\tau)$ reaches its minimum, namely $\sigma_{Y_I}^2(\tau) = \sigma_{y'_I}^2(\tau)$, and it can be deduced to:

$$K_I(\tau) = \text{cov}_A[y(\tau), x_I(\tau)]/\sigma_{x_I}^2(\tau) \tag{8}$$

Then we obtain the expression of FSC from the statistical perspective, which is equal to the ratio of ACOV between $y(\tau)$ and $x_I(\tau)$ to ADEV of $x_I(\tau)$. Here normal covariance and deviation are replaced by ACOV and ADEV, due to characteristics of time and frequency noise, which have several kinds of power laws with $\tau$, so are some NIVs. From Eq.(8), we obtain a curve of $K_I(\tau)$ with $\tau$ firstly, then calculate to get an exact value from the curve, by evaluating the points of curve and their confidences. If both $\{\Delta y_j(\tau)\}$ and $\{\Delta x_{I,j}(\tau)\}$ satisfy the normal distribution with expected values of 0, we can get:

$$\frac{\sigma_{K_I}^2}{K_I^2} \xrightarrow{M \to \infty} \frac{1}{M}\left(0.47\frac{\sigma_y^2}{\sigma_{y_I}^2} + \frac{1}{K_M}\right) \tag{9}$$

where $\sigma_{y_I}^2 = k_I^2 \cdot \sigma_{x_I}^2$ is the contribution of $x_I$ to $\sigma_y^2$. $K_M$ is the coefficient related to power law of $\sigma_{x_I}^2(\tau)$ which is directly proportional to $\tau^{-1}$, $\tau^0$ and $\tau$ for white frequency noise (WFN), flicker frequency noise (FFN) and random walk noise (RWN), respectively[21]. For WFN, FFN and RWN, $K_M \approx 0.87, 0.77, 0.75$, respectively. In the following discussion, $K_M$ is set to be 0.75, corresponding to the minimum value of $K_M$ for large uncertainty. Eq.(9) results from several approximations, which are not fully satisfied, especially the approximation of $M \to \infty$, these will lead to the errors of $\sigma_{K_I}^2$. Since there is not better expression, Eq.(9) can be applied for all (in fact, only data of $M \geq 1000$ is discussed below) $M$ and all kinds of noise. Nevertheless, Eq.(9) maps two most important characteristics of $\sigma_{K_I}^2$, being inversely proportional to $M$ and approximate inverse ratio to weight of noise of $x_I$ in total uncertainty ($\frac{\sigma_{y_I}^2}{\sigma_y^2}$).

In experiments, the NSC method requires measuring $\{y_j(\tau)\}$ and $\{x_{I,j}(\tau)\}$ precisely and synchronously. Usually, $y_j(\tau)$ can't be measured directly, but replaced by $y_{com\_j}(\tau)$, which is comparing fractional frequency of AFS with other frequency reference $y_{ref\_j}(\tau)$ and satisfied $y_{com\_j}(\tau)=y_j(\tau)-y_{ref\_j}(\tau)$. Here $y_{ref\_j}(\tau)$ is considered to be independent of $y_j(\tau)$. According to Eq.(9), the test accuracy of $K_I$ will be decreased due to the introduction of reference noise. The measurement result of $x_I$ should be the value of $x_{I,j}(\tau)$ which affects quantum medium when they interact with microwave or optical field at exact time interval from $(j-1)\tau$ to $j\tau$. In fact, $x_{I,j}(\tau)$ has its $u_B$ and asynchrony error too, the latter comes from indirectly measurement of some NIVs, such as temperature measurements of interacting zone for either AFFSs or OFSs. Three most common forms of error of $x_{I,j}(\tau)$ can be expressed as: $x_I(\tau)=x_{I\_r}(\tau)+x_{I\_B}$, $x_{I\_r,j}(\tau)=x_{I,j}(\tau+\tau_{dly})$, $x_{I\_r,j}(\tau)=\frac{1}{\tau_{int}}\int_{\tau-\tau_{int}/2}^{\tau+\tau_{int}/2}x_{I,j}(\tau')d\tau'$ respectively, where $x_{I\_r}(\tau)$ and $x_I(\tau)$ are the real and measurement value of $x_I$ at corresponding $\tau$ time interval, respectively. The first expression describes the effect of $u_B$ of $x_I$, which will lead to the decrease of $K_I$, fitting $K_I=\frac{\sigma_{x_{I\_r}}^2(\tau)}{\sigma_{x_I}^2(\tau)}k_I$. The second describes that $x_{I\_r}(\tau)$ have time delay of $\tau_{dly}$ to $x_I(\tau)$, and the third represents $x_{I\_r}(\tau)$ is integral mean of $x_I(\tau)$ from $\tau-\tau_{int}/2$ to $\tau+\tau_{int}/2$. Considering the value $K_I(\tau)/k_I$ is 1 in the time overlap part of $y(\tau)$ and $x_{I\_r}(\tau)$, and is equal to 0 in the time non-overlap part, the modified expression of $K_I(\tau)$ is described as:

$$K_{I\_dly}(\tau)=\begin{cases}0 & \tau<|\tau_{dly}|/2 \\ (|\tau_{dly}|/2\tau-1)k_I & |\tau_{dly}|/2\leq\tau<|\tau_{dly}| \\ (1-3|\tau_{dly}|/2\tau)k_I & \tau\geq|\tau_{dly}|\end{cases} \qquad(10)$$

$$K_{I\_int}(\tau)=\begin{cases}0 & \tau<\frac{\tau_{int}}{4} \\ \frac{(\tau_{int}-4\tau)^2}{8\tau\cdot\tau_{int}}k_I & \frac{\tau_{int}}{4}\leq\tau<\frac{\tau_{int}}{2} \\ \left(1-\frac{3}{8}\frac{\tau_{int}}{\tau}\right)\cdot k_I & \tau\geq\frac{\tau_{int}}{2}\end{cases} \qquad(11)$$

Where $K_{I\_dly}(\tau)$ and $K_{I\_int}(\tau)$ represent the variation of $K_I(\tau)$ for time-delay and time integral mean in the measurement of $x_I$, respectively, shown as dotted blue and dashed pink lines in Fig. 2. The asynchrony of $x_I$ makes $y(\tau)$ and $x_I(\tau)$ correlated partially. When $\tau \leq \tau_{dly}$ or $\tau \leq \tau_{int}$, $K_{I\_dly}(\tau)$ or $K_{I\_int}(\tau)$ are affected by $\tau_{dly}$ or $\tau_{int}$ obviously. With the increase of $\tau$, proportion of correlation part tends to 1, then the curves trend to $K_I(\tau)=k_I$. Therefore, the asynchrony of $x_I$ can be distinguished from the curve of $K_I(\tau)$, which exists in $u_B$ evaluations of many effects. And the asynchrony of $x_I$ can be amend by replacing $\{x_{I,j}(\tau)\}$ with a modified sequence $\{x'_{I,j}(\tau)\}$, then a modified $K_I(\tau)$ can be obtained, where $x'_{I,j}(\tau) = x_{I,j}(\tau+\tau_{dly})$ or $x'_{I,j}(\tau) = \frac{1}{\tau_{int}} \int_{\tau-\tau_{int}/2}^{\tau+\tau_{int}/2} x_{I,j}(\tau')d\tau'$.

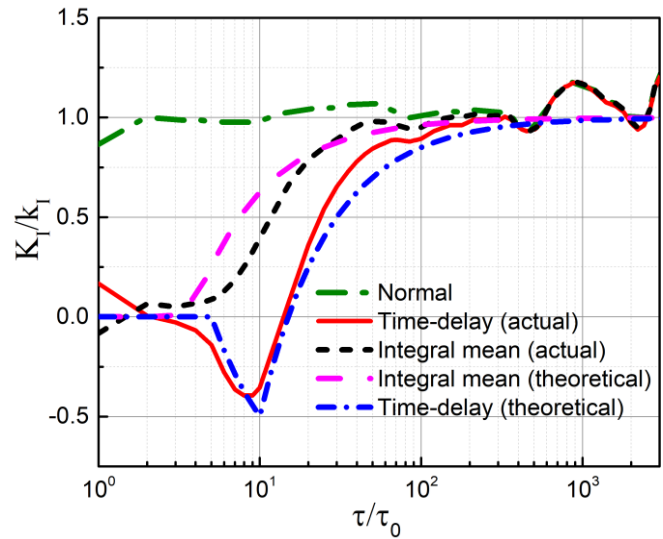

Figure 2 (color online). The typical curves of $K_I(\tau)$ versus the test and actual values of $x_I(\tau)$ having time-delay or integral mean. Here both $\tau_{dly}$ and $\tau_{int}$ are equal to $10\tau_0$. The dash-dotted olive line represents the case without asynchrony, and the dashed pink and short dash-dotted blue, short-dashed black and solid red lines describe the theoretical and actual values in the presence of time integral mean and time-delay, respectively.

Numerical Simulation

The NSC method is validated by numerical simulation. As shown in Eq.(9), the precision of FSC is related to the noise types of $y(\tau)$ and $x_I(\tau)$, so three kinds of typical data of $y(\tau)$ and $x_I(\tau)$ are gave to verify. The first data is real data coming from comparing experiments of $^{87}$Rb AFFS and H-maser, which usually consists of many kinds of noise. The second data is ideal single noise of WFN, FFN or RWN created by computer, and the last data is asynchrony data of $y(\tau)$ and $x_I(\tau)$. The results are shown in Fig.3, Fig4 and Fig.5, respectively.

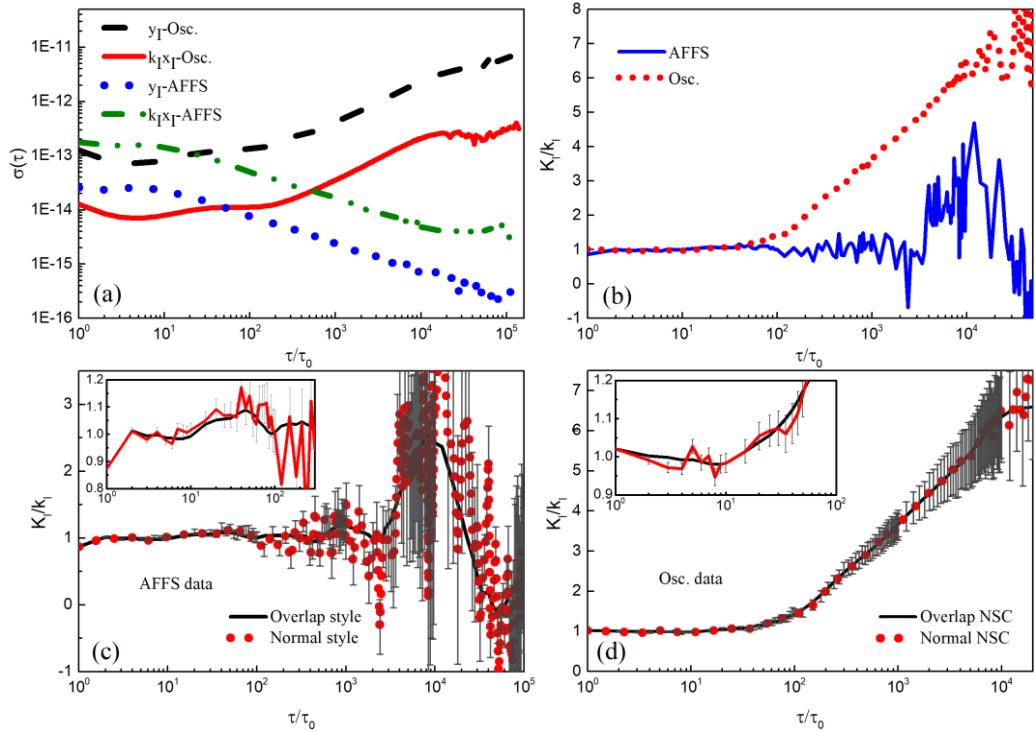

Figure 3 (color online). Validated results of the NSC method by numerical simulation. (a) the curves of Allan deviation for the sequences of $\{y_j(\tau_0)\}$ and $\{k_I \cdot x_{I,j}(\tau_0)\}$ of Osc. and AFFS, represented by dashed black, solid red, dash-dotted olive and dotted blue lines, respectively. Here the proportion of $\frac{\sigma_{y_I}^2}{\sigma_y^2}$ is about 0.01. (b) $K_I(\tau)$ curves of two NS models: AFFS and Osc., shown as solid blue and dotted red lines, respectively. The error increases with the decrease of $M$ ($\propto 1/\tau$), but the rates of growth are different. (c) and (d) are curves comparing of AFFS and Osc. models, obtained by overlap and normal style, shown as solid black and dotted red lines, respectively.

Fig.3 plots the results of numerical simulation whose noise sequences come from the past comparing data of $^{87}$Rb AFFS and its local oscillator (Osc.) with reference to a hydrogen maser (VCH- 1003A) through the combinations of "AFFS + AFFS" and "Osc. + Osc." with $M_0 = 6.5 \times 10^5$ and $M_0 = 6.9 \times 10^5$, respectively. $M_0$ is the sample number of $\{y_j(\tau_0)\}$ and $\{k_I \cdot x_{I,j}(\tau_0)\}$ sequences, $\tau_0$ is the running period of the AFS, $\{k_I \cdot x_{I,j}(\tau_0)\}$ comes from another comparing data $\{y_{2,j}(\tau_0)\}$, written as $\{0.11 * y_{2,j}(\tau_0)\}$, which means the proportion of $\frac{\sigma_{y_I}^2}{\sigma_y^2}$ is about 0.01 at all $\tau$ coordinate interval and the expected value of $K_I(\tau)$ is 0.11. The two sequences $\{y_{1,j}(\tau_0)\}$ and $\{y_{2,j}(\tau_0)\}$ are independent of each other since the time interval between their data records is more than one month. The sequence of $\{y_j(\tau_0)\}$ is created by the expression of $y_j(\tau_0)=y_{1,j}(\tau_0)+0.11*y_{2,j}(\tau_0)$. Fig.3 shows that the error of $K_I(\tau)$ is related to noise type and $M$. As a theoretical expectation, the fractional error of $K_I(\tau)/k_I$ is less than 0.1 when $M \geq 1000$ ($\tau/\tau_0 \leq 650$) for AFFS data and $M \geq 1.7 \times 10^4$ ($\tau/\tau_0 \leq 40$) for Osc. data. In order to use the data more efficiently, an overlap data processing style is applied to create modified $K_I(\tau)$ curves[①], which only reduces the discreteness of the adjacent points on the curve and makes the curve smoother, as shown in Fig. 3(c) and (d). The shape of curve is coincident with that of normal style. Error bars of $K_I(\tau)$ are also shown in Fig.3(c) and (d). $M$ is an approximation expression of enlarging $\sigma_{K_I}$, real $M$ is replaced by *edf*, the equivalent degree of freedom in overlap style[21], with complex expressions for different noise. The $K_I(\tau)$ curves with overlap style are more

---

① It means writing $\sigma_y^2(\tau)$ in the form of overlap ADEV, and overlap ADEV and ACOV are written as:

$$\sigma_{y-ov.ADEV}^2(\tau) = \frac{1}{2m^2(M-2m+1)} \sum_{j=1}^{M-2m+1} \left\{ \sum_{i=j}^{j+m-1} \left[ y_{i+1}(\tau_0) - y_i(\tau_0) \right] \right\}^2 \text{ and}$$

$$\text{cov}_{ov.ADEV}(a,b,m \cdot \tau_0) = \frac{1}{2m^2(M-2m+1)} \times \sum_{j=1}^{M-2m+1} \left\{ \sum_{i=j}^{j+m-1} \left[ a_{i+1}(\tau_0) - a_i(\tau_0) \right] \right\} \cdot \left\{ \sum_{i=j}^{j+m-1} \left[ b_{i+1}(\tau_0) - b_i(\tau_0) \right] \right\}$$

convenient to evaluate $K_I$, so in the following discussion, overlap style is used to process $K_I(\tau)$.

Comparing Fig.3(c) and Fig.3(d), we can see that $K_I(\tau)$ curve of Osc. data deviates from the real value faster than theoretical expectation with the increase of $\tau$, it is beyond error bar when $\tau \geq 100\tau_0$ (while it is coincident for $K_I(\tau)$ curve of AFFS data even $\tau \geq 10^4 \tau_0$ ). The difference between two curves should come from the difference of noise types, mainly FFN and RWN for Osc. and WFN for AFFS. Numeral simulations of single type of noise are used to verificate the conjecture. The noise sequences are created by Stable 32 software[22], $M_0 = 2 \times 10^6$ for every sequence, which is also divided into four equal parts of $M_0' = 5 \times 10^5$ to observe the statistical characteristics of corresponding curves. The results are shown in Fig.4, WFN and FFN are consistent with theory when $M \geq 1000$, corresponding to $\tau \leq 2000\tau_0$ or $\tau \leq 500\tau_0$. For their different $M_0$ and ratio of uncertainty of $\frac{\sigma_{y_I}^2}{\sigma_y^2}=0.05$, we can obtain the value of $K_I$ with uncertainty better than 10% when $M \geq 10^4$. For RWN, the NSC method is also effective, but requires far larger $M$ than WFN and FFN sequences (about 10~100 times) in order to obtain the same precision of $K_I$ as WFN and FFN. And Eq.(9) does not effectively describe $\sigma_{K_I}$ any more, we should give a more universal expression for fitting the condition of RWN, but thinking of Eq.(8) is right, so modified Eq.(9) must enlarge the expectation of $\sigma_{K_I}$ RWN for coinciding the experiments or NS result, which is useless for increasing the precision of $K_I$. A better way is to search improved algorithm for decreasing $\sigma_{K_I}$ of RWN. A noise statistical correlation difference method (NSC-D) is proposed that replacing $\{y_j(\tau_0)\}$ and $\{x_{I,j}(\tau_0)\}$ with their differences, $\{\Delta y_j(\tau_0)=y_{j+1}(\tau_0)-y_j(\tau_0)\}$ and $\{\Delta x_{I,j}(\tau_0)=x_{I,j+1}(\tau_0)-x_{I,j}(\tau_0)\}$ in Eq.(8) and obtaining the modified $K_I(\tau)$. Due to eliminating relation of elements of RWN sequence, an improved $K_I(\tau)$ curve is obtained, as shown in Fig.4(d). The result achieves the expected improvement and the curve is limited in error bar when $\tau \leq 300\tau_0$. The range is an order of magnitude larger than normal NSC, but NSC-D also has some shortcomings, for example, it is invalid once the synchronization is not satisfied. Acting as an auxiliary tool, NSC-D will be researched in the future.

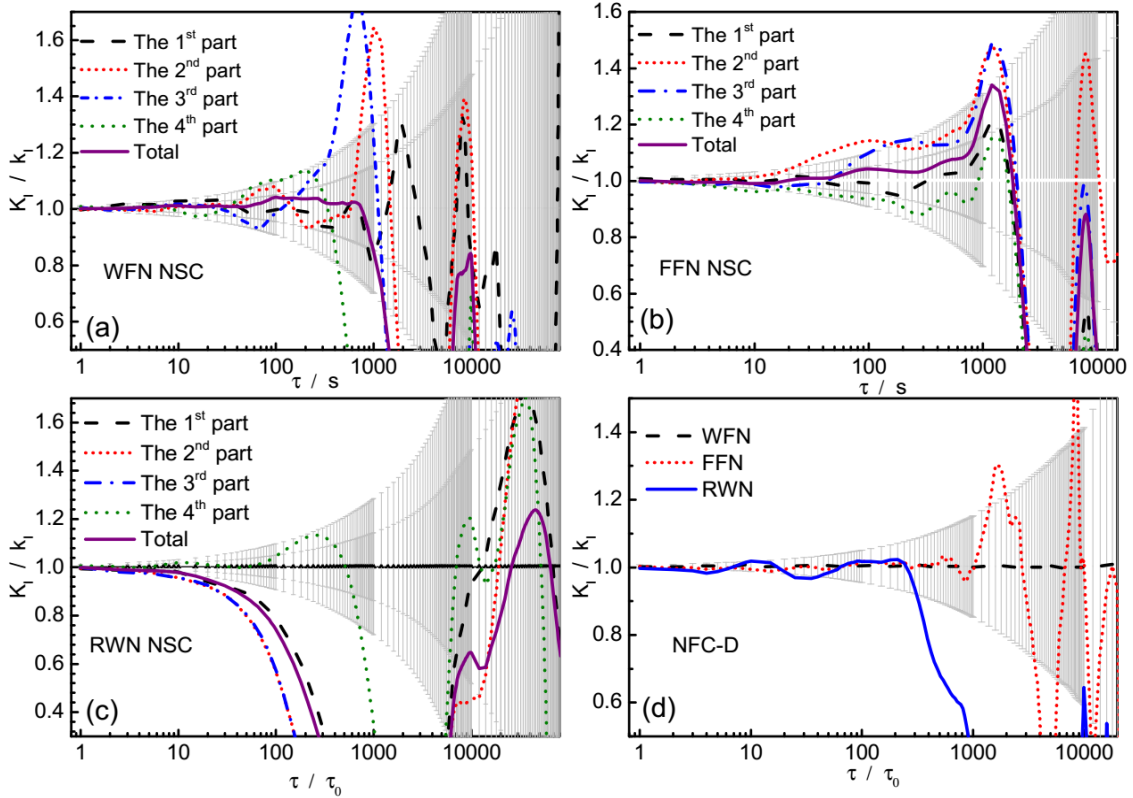

Figure 4 (color online). Validated results of the NSC method for the three types single noise models by numerical simulation, shown as solid purple line and $\frac{\sigma_{y_I}^2}{\sigma_y^2}=0.05$. (a) WFN. (b) FFN. (c) RWN. (d) results of a modified NSC method (NSC-D) which replaces the noise data with their difference between adjacent data. Here $M_0 = 2\times 10^6$ for every noise sequences and which is also divided into four equal parts for comparing the results of these four pieces of $K_I(\tau)$ curves with $M_0' = 5\times 10^5$, in (a), (b) and (c) four parts are marked as dash-dotted black, short-dotted red, dashed blue and dotted olive, respectively. Two error bar ranges in (a), (b) and (c) is corresponding of $\sigma_{K_I}$ of $M_0 = 2\times 10^6$ and $M_0' = 5\times 10^5$, respectively.

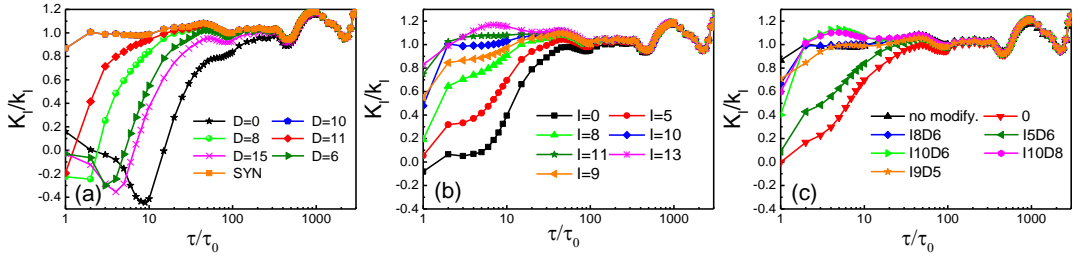

Figure 5 (color online). NS validated results of the NSC method for compensating asynchrony. (a) time-delay compensation, $\tau_{dly}=10\tau_0$. (b) time integral compensation, $\tau_{int}=10\tau_0$. (c) the results of compensation of time-delay and integral.

We have also validated the capacity of distinguishing asynchrony of the NSC method by numeral simulation. The noise sequences are created from same data of $^{87}$Rb AFFS, replacing $\{0.11^*y_{2,j}(\tau_0)\}$ with its modified expressions of time-delay or integral mean. The results of numeral simulation are shown in Fig.2 (b), the curves of $K_I(\tau)$ are well coincident with the theory, which has validated the capability of NSC measuring asynchrony of $y(\tau)$ and $x_I(\tau)$. While for the NSC method, the asynchrony will lead to additional error of $K_I(\tau)$ at $\tau \sim \tau_0$ interval, which is the highest accuracy end of the curve due to largest sample numbers if there is no asynchrony. So asynchrony compensation is used to restrain the effect asynchrony by constructing the sequence of $\{x'_{I,j}(\tau)\}$, the results of numeral simulation are shown in Fig.5. Seen from Fig.5, time-delay and time integral can be compensated but the latter with lower accuracy. There is almost no time-delay or integral alone, and in most cases both exist at the same time. Moreover, we can obtain the optimal compensation coefficients with D=10 for time-delay compensation in Fig.5(a), I=10 for integral mean compensation in Fig.5 (b) and D=6 and I=8 for time-delay and integral mean compensation at the same time in Fig.5 (c), these values are the real values of the asynchrony. The mainly asynchrony can be compensated following the modified $K_I(\tau)$ curves, and after compensation, the error of $K_I(\tau)$ nearby $\tau_0$ will decrease remarkably.

In principle, $K_I(\tau)$ should be a $\tau$-independent constant rather than a $\tau$-dependent curve. However, as shown above, NSC is a full statistical method, its curve is not smooth but with larger drifts at two ends. The reasons lead to the drift at two ends are different, at the side of $\tau \gg \tau_0$, it is main statistical error due to reducing of samples number, which is a mathematical problem, while at the end of $\tau \sim \tau_0$, it is because the indirectness of time and space on detecting $x_I(\tau)$, which is a physical problem. In fact, the curve part nearby $\tau \sim \tau_0$ gives definite $K_I(\tau)$ resulting from the indirect measurement of $x_I(\tau)$. So this drift nearby $\tau \sim \tau_0$, which can be seen as a criterion for measurement of $x_I(\tau)$ give us a reference for evaluating and improving the measurement of $x_I(\tau)$.

Here, we consider $K_I(\tau)$ is the result cannot be improved by physical way, and look for unified $K_I(\tau)$ from the curve. There is not a standard method for evaluating ultimate value of $K_I$ from $K_I(\tau)$ curve, we point out one that considers the shape of $K_I(\tau)$ curve with the error bars of points, described as following: (1). Select the range of $\tau$ covering an interval of more than one order of magnitude, where the error bars of points are minimum, and value of every point is fall in the errors bars of middle points. (2). Calculate average $\overline{K_I}$ and standard variance $\sigma_{\overline{K_I}}$ of points on the curve in the interval, (3). $\overline{K_I}$ means the ultimate value of $K_I$ and its uncertainty is expressed as square root of sum of squares $\sigma_{\overline{K_I}}$ and the maximum error bar of the points. The results are shown in table 1. From the table, we know the fractional precision of $K_I$ based on above data is about $1\% \sim 5\%$, and its errors and uncertainties are consistent. In theory, the NSC method can improve the precision continuously by increasing experiment time

(enlarging $M_0$). But considering $\sigma_{K_I} \propto 1/\sqrt{M_0}$, it is difficult to further improve its precision when $M_0$ reaches $10^5 \sim 10^6$. So for the NSC method, the measurement precision of $K_I$ is easy to reach $1\% \sim 10\%$, but difficult to obtain higher precision.

Table 1. The ultimate value of $K_I$ coming from the curve of $K_I(\tau)$ and its error bar.

| NOISE | METHOD | SAMPLING INTERVAL | $\overline{K_I}$ | $\sigma_{\overline{K_I}}$ | $\sigma_{K_I\_MAX}(\tau)$ | TOTAL $\sigma_{K_I}$ |
|---|---|---|---|---|---|---|
| OSA | NSC | 1-20 | 1.002 | 0.02 | 0.03 | 0.04 |
| AFFS | NSC | 2-30 | 1.01 | 0.02 | 0.04 | 0.04 |
| WFN | NSC | 1-50 | 1.007 | 0.005 | 0.004 | 0.007 |
| FFN | NSC | 1-20 | 0.996 | 0.005 | 0.004 | 0.007 |
| RWN | NSC | 1-15 | 0.996 | 0.04 | 0.02 | 0.05 |
| RWN | NSC-D | 1-40 | 1.003 | 0.02 | 0.01 | 0.02 |
| AFFS (I) | NSC (I=11) | 3-60 | 1.04 | 0.02 | 0.03 | 0.04 |
| AFFS (D&I) | NSC | 2-20 | 1.005 | 0.06 | 0.014 | 0.06 |

## Demonstration experiments

Experiments are carried out with $^{87}$Rb AFFS device to prove the NSC method. Fig. 1 (c) demonstrates the setup of the experiments. An additional noise is introduced to the AFFS for the experiments by adding a current noise $\delta I$ on the solenoid coil which provides uniform magnetic field for Ramsey interaction. $\delta I$ brings the noise of magnetic intensity, and affects the output of AFFS by second-order Zeeman shift, fitting $y_{j,2-Z}(\tau) = a_{2-Z} \cdot [\overline{I} + \delta I_j(\tau)]^2$, where $a_{2-Z}$ is the simplified proportionality coefficient of second-order Zeeman shift, and $\delta y_{j,2-Z}(\tau) = 2a_{2-Z}\overline{I} \cdot \delta I_j(\tau)$. The aim of NSC is to evaluate $k_{2-Z}(I) = 2a_{2-Z}\overline{I}$. The setup is similar to that of our another work[23], whose object is to decrease the noise when $k_{2-Z}(I)$ is known. In experiments, $y_j(\tau)$ and

$\delta I_j(\tau)$ are replaced by $y_{com\_j}(\tau)$ and $\delta V_j(\tau)=\delta I_j(\tau)\cdot R$, respectively. It is easier to obtain by measuring the voltage of a precision sampling resistor $R$, in series connection with the coil circuit, so $k_{2-Z}(I)$ is turn to $k_{2-Z}(V)$, written as $k_{2-Z}$. Fitting $k_{2-Z}=k_{2-Z}(I)/R$ and $\delta y_{j,2-Z}(\tau)=k_{2-Z}\cdot \delta V_j(\tau)$, which can be calibrated by other method, such as Zeeman shift of the other magnetic sensitive transitions, the result of $k_{2-Z}=6.47\times 10^{-14}\text{V}^{-1}$ is obtained. Two noise sequences of $\delta V_j(\tau)$ are applied in the experiments, one is the same sequence used in Ref.23, the other is random noise created by computer, which is an order of magnitude lower than the frequency stability of the AFFS. The results are shown in Fig.6, the curves show obvious asynchrony, it is because that $\delta V_j(\tau)$ affects the reference frequency of quantum system while $y_j(\tau)$ is the fractional frequency of local oscillator. The process from phase discrimination to feedback leads to $y_j(\tau)$ being about a periodic time-delay relative to $\delta V_j(\tau)$, and the transfer function of the error signal and the proportional integral differential (PID) feedback parameter make asynchrony of $y_j(\tau)$ and $\delta V_j(\tau)$ much complex, as shown in Fig.6 (b) and (c). When NSC is used, the curves become flatter than the original ones clearly, and obviously, we cannot compensate the asynchrony completely. Based on the improved results, $\overline{K_I}$ is measured with values of $6.99\times 10^{-14}\text{V}^{-1}$ and $6.83\times 10^{-14}\text{V}^{-1}$, and relative uncertainties of 0.11 (0.08/0.08) for $\sigma_{\overline{K_I}}/\sigma_{K_I\_MAX}(\tau)$) and 0.11 (0.1/0.03), respectively, the result is consistent with $6.47\times 10^{-14}\text{V}^{-1}$ in $1\sigma$ confidence interval, the value is obtained by measuring the curve of magnetic field with voltage. It requires further studies on why the results of two NSC experiments are larger than the value of other method.

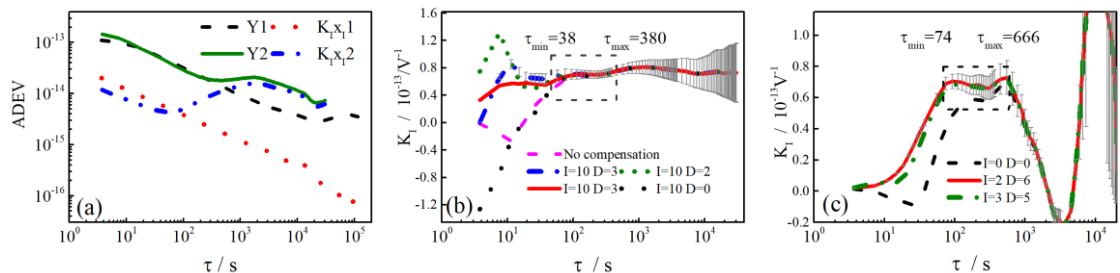

Figure 6 (color online). The results of demonstration experiments. (a) the stability curves of two experiments. (b) and (c) are the curves of $k_{2\_Z}$ of two $\delta V_j(\tau)$ noise sequences experiments. The none-compensation curves in (b) and (c) show obvious asynchrony of $\delta V_j(\tau)$ measurement, are result from complex time-dependent relation of AFFS lock-in. From the curves, we can see that $\overline{K_I}$ are $6.99 \times 10^{-14} \text{V}^{-1}$ and $6.83 \times 10^{-14} \text{V}^{-1}$, $\sigma_{K_I\_total} / \overline{K_I}$ are 0.11 (0.08/0.08) and 0.11 (0.1/0.03) for two experiments, respectively.

## Discussion and prospects

Different from the standard method that $u_B$ and $u_A$ are independent and total uncertainty fits $u_{total}^2 = u_A^2 + u_B^2$, the NSC method considers the contribution of $x_i(\tau)$ to $u_A$, expressed as Eq.(2) and is similar to $u_B$. By bringing $\{y_j\}$ and $\{x_{i,j}\}$ into the Eq.(7), we can obtain $k_i$ by solving the minimum value of $\sigma_{Y_I}(\tau)$. The difference of principles between the two methods leads to the differences of practice: (1). The standard method requires specialized experiment for measuring $k_i$ while NSC needs to record the values of $x_i$ when the device is running, therefore NSC is more suitable for continuous running devices. (2). NSC has capability of parallel processing, which can synchronously evaluate every $k_i$ as long as its NIV can be measured precisely in real-time. (3). The standard method evaluates $u_B$ with different effects while NSC evaluates $u_B$ with different NIVs. (4). NSC is more tolerant to the noise. The only way to reduce $u_B$ with the normal method is to suppress noise, while NSC only need to detect the noise precisely, and then decrease $u_A$ and $u_B$ by some ways, such as real-time noise distinguish (RTND)[23].

There are three hypothesizes in deducing of $K_I(\tau)$: $M$ tends to be infinite,

different $x_i(\tau)$ are independent of each other, and the first-order Taylor approximation is valid. We have discussed the first condition of $M$ act as a definite finite value. For the other two hypothesizes, they are pointed out only for simplification. If different $x_i(\tau)$ are correlated, and if the expression of $y_i = f(x_i)$ is nonlinearity, we just need to write expressions with more parameters and more covariance, then we can solve as above deducing way.

NSC method has many applications. It's a way to test the standard method,. For example, we can validate the result of BBR shift evaluation of optical clock by the NSC method. For NSC records data when device is running, it gives a method of monitoring $u_B$, which is important for continuously running devices such as AFSs and similar setups, and cannot be realized by other method. NSC is expected to help us to find some unknown effects hidden in total $u_B$, and as above shown, whose FSCs can be detected even if its contribution is far less than $\sigma(\tau)$. And the results of the method also show a capability that can predict the contribution of $x_i$ for long-term stability by evaluating $k_i$ from its short-term interval data. And the NSC method can find asynchrony of the measurement of $\{x_{i,j}\}$ by the curve of $K_i(\tau)$. Above all, NSC is a useful method for uncertainty, which not only can test the standard method, but also can extract more information of $u_B$ evaluations out of the standard method.

Form of Eq.(8) is the same as the expression of slope by linear fitting of LSM. Eq.(8) is deduced from the condition that different noises are independent of each other while LSM is directly pointed out as an effective fitting method. Here, NSC is used to deal with the correlation of variables in form of noise, while LSM is used to fit lines or curves with obvious signal to noise, the difference can be looked on different ranges of SNR (SNR <<1 VS. SNR >1) while the method of minimum sum of squares is equally effective. At the same time, NSC give a physical explanation for the validity of LSM is given as follows: when input quantity is independent of the noise that affects output quantity, the ratio coefficient of input to output quantity satisfies Eq.(8), which is so-called as linear fitting of LSM. By comparing NSC and LSM, we can obtain more interesting results, on one hand, referring from LSM, we can use NSC to deal with complex correlations between $y$ and $x_I$, on the other hand, NSC is helpful for understanding relationships along different noises, which can be used to improve the validity of LSM.

In conclusion, we have proposed a NSC method for evaluating FSCs and $u_B$, which is based on the real-time relationship of signal output of AFS and its environmental NIVs, giving the statistical expression of FSC as $K_I(\tau) = \text{cov}_A[y(\tau), x_I(\tau)]/\sigma_{x_I}^2(\tau)$, then evaluating $u_B$ by $u_B^2 = \sum k_i^2 \sigma_{x_i}^2$. The method has been demonstrated by NS and experiments, the value of $K_I$ is measured with uncertainty of 1%~10% when $M \approx 10^5 \sim 10^6$ and $\sigma_{y_I}^2/\sigma_y^2$ is about 10%. Next time, we will use the NSC method to evaluate collision shift[5, 24] of AFFS, and we believe it is effective for other effects of other AFSs, such as blackbody radiation shift[13, 14, 15], optical frequency shift[16, 17, 18] in OFS. The method not only can evaluate main $u_B$ of most accuracy AFSs, but also is benefit to judge what and how do NIVs affect the stability of those compact AFSs and similar devices, as well as extends to other fields of universal measurement. Furthermore, we have given a method to decrease the uncertainty contribution of the effect in our another work[23]. And it should be pointed out, the theory of NSC is still preliminary, and there is a large error between theoretical expectation and the results of NS or experiments, more rigorous derivation of NSC is needed as well as more efficient algorithms are required for data processing.

We thank our colleague Tang Li and professor Jinming Liu of east china normal university for useful discussion. And the work is supported by the Strategic Priority Research Program of the Chinese Academy of Sciences. Grant NO.XDB21030200

1. R. Wynands and S. Weyers, Atomic fountain clocks, Metrologia **42**, 64 (2005).

2. J. Guena, S. Weyers, M. Abgrall, C. Grebing, V. Gerginov, P. Rosenbusch, S. Bize, B. Lipphardt, H. Denker, N. Quintin, S. M. F. Raupach, D. Nicolodi, F. Stefani, N. Chiodo, S. Koke, A. Kuhl, F. Wiotte, F. Meynadier, E. Camisard, C. Chardonnet, Y. Le Coq, M. Lours, G. Santarelli, A. Amy-Klein, R. Le Targat, O. Lopez, P. E. Pottie and G. Grosche, First international comparison of fountain primary frequency standards via a long distance optical fiber link Metrologia **54**, 348 (2017)

3. S. Peil, J. L. Hanssen, T. B. Swanson, J. Taylor and C. R. Ekstrom. Evaluation of long term performance of continuously running atomic fountains, Metrologia **51**, 263 (2014).

4. A. D. Ludlow, M. M. Boyd, J. Ye, E. Peik and P. O. Schmidt, Optical atomic clocks, Rev. Mod. Phys. **87**, 637 (2015).

5. K. Gibble and S. Chu, Laser-Cooled Cs Frequency standard and a measurement of the frequency-shift due to ultracold collision. Phys. Rev. Lett. **70,** 1771 (1993).

6. Y. Sortais, S. Bize, C. Nicolas, A. Clairon, C. Salomon, C. Williams, Cold Collision Frequency Shifts in a 87Rb Atomic Fountain, Phys. Rev. Lett. **85** 3117 (2000).

7. R. Li and K. Gibble, Evaluating and minimizing distributed cavity phase errors in atomic clocks, Metrologia **47** 534 (2010).

8. J. Guéna, R. Li, K. Gibble, S. Bize and A. Clairon, Evaluation of Doppler Shifts to Improve the Accuracy of Primary Atomic Fountain Clocks, Phys. Rev. Lett. **106** 130801 (2011).

9. R Li, K Gibble and K Szymaniec, Improved accuracy of the NPL-CsF2 primary frequency standard: evaluation of distributed cavity phase and microwave lensing frequency shifts, Metrologia **48** 283 (2011).

10. S Weyers, V Gerginov, N Nemitz, R. Li and K. Gibble, Distributed cavity phase frequency shifts of the caesium fountain PTB-CSF2, Metrologia **49** 82 (2012).

11. S. Weyers, U. Huebner, R. Schroeder, C. Tamm, and A. Bauch, Uncertainty evaluation of the atomic caesium fountain CSF1 of the PTB, Metrologia **38** 343 (2001).

12. K. Szymaniec, S. E. Park, G. Marra and W. Chałupczak, *First accuracy evaluation of the NPL-CsF2 primary frequency standard,* Metrologia **47** 363 (2010).

13. S. G. Porsev and A. Derevianko, Multipolar theory of blackbody radiation shift of atomic energy levels and its implications for optical lattice clocks, Phys. Rev. A **74** 020502 (2006)

14. I. Ushijima, M. Takamoto, M. Das, T. Ohkubo and H. Katori, Cryogenic optical lattice clocks, Nature Photon **9** 185 (2015)

15. T. L. Nicholson, S. L. Campbell, R. B. Hutson, G. E. Marti, B. J. Bloom, R. L. McNally, W. Zhang, M. D. Barrett, M. S. Safronova, G. F. Strouse, W. L. Tew, and J. Ye, Systematic evaluation of an atomic clock at 2 × 10−18 total uncertainty, Nature Communications **6** 6896 (2015)

16. T. Akatsuka, M. Takamoto and H. Katori, Three-dimensional optical lattice clock with bosonic Sr-88 atoms,Phys.

Rev. A **81** 023402 (2010)

[17] S. Falke, H. Schnatz, J. S. R. V. Winfred, T. Middelmann, S. Vogt, S. Weyers, B. Lipphardt , G.Grosche, F. Riehle, U. Sterr and C. Lisdat, The 87Sr optical frequency standard at PTB. Metrologia **48** 399 (2011).

[18] C. Y. Park, D. H. Yu, W. K. Lee, S. E. Park, E. B. Kim, S. K. Lee, J. W. Cho, T. H. Yoon, J. Mun, S. J. Park, T. Y. Kwon, S. B. Lee, absolute frequency measurement of 1S0(F = 1/2)–3P0(F = 1/2) transition of 171Yb atoms in a one-dimensional optical lattice at KRISS. Metrologia **50** 119 (2013)

[19] JCGM 100:2008, Evaluation of measurement data-Guide to the expression of uncertainty in measurement, https://www.bipm.org/utils/common/documents/jcgm/JCGM_100_2008_E.pdf

[20] F. Riehle, , 2004, Frequency Standards: Basics and Applications, (Wiley-VCH, Weinheim).

[21] W. J. Riley, 2008, Handbook of Frequency, Stability Analysis, NIST, Special Publication 1065 (NIST, Boulder).

[22] http://www.wriley.com/

[23] R. C. Dong, J. D. Lin, R. Wei , W. L. Wang, F. Zou, Y. B. Du, T. T. Chen and Y. Z. Wang, Decreasing the uncertainty of atomic clocks via real-time noise distinguish, Chinese Optics Letters **15** 050201 (2017)

[24] Y. Sortai s, S. Bize, C. Nicolas, A. Clairon, C. Salomon and C. Williams, Cold Collision Frequency Shifts in a 87Rb Atomic Fountain. Phys. Rev. Lett. **85** 3117 (2000).